\documentclass[letterpaper, 10 pt, conference]{ieeeconf}  % Comment this line out if you need a4paper

\IEEEoverridecommandlockouts                              % This command is only needed if 
\usepackage{amsmath}
\usepackage{amssymb}
\usepackage{comment}
\usepackage{multirow}
\usepackage{cite} % to get multiple citations as [11]-[13] instead of [11], [12], [13]
\usepackage{lipsum}% http://ctan.org/pkg/lipsum
\usepackage{graphicx}% http://ctan.org/pkg/graphicx
\usepackage{mathtools}
\usepackage{url}
\usepackage{hyperref}
\usepackage{etoolbox}
\allowdisplaybreaks
\DeclarePairedDelimiter{\ceil}{\lceil}{\rceil}

\newcommand{\bs}{\mathbf{s}}
\newcommand{\ba}{\mathbf{a}}

\newcommand{\bA}{\mathbf{A}}
\newcommand{\bD}{\mathbf{D}}
\newcommand{\bW}{\mathbf{W}}
\newcommand{\bI}{\mathbf{I}}

\newtoggle{extended}
\toggletrue{extended}
\usepackage[dvipsnames]{xcolor}
\newcommand{\highest}[1]{\textcolor{Maroon}{{#1}}}%

\title{\LARGE \bf
Real-time Control of Electric Autonomous Mobility-on-Demand Systems via Graph Reinforcement Learning 
}

\iftoggle{extended}{

}{
    \renewcommand{\baselinestretch}{1}
}

\author{Aaryan Singhal$^{1}$, Daniele Gammelli$^{1}$, Justin Luke$^{1}$, Karthik Gopalakrishnan$^{1}$, Dominik Helmreich$^{2}$ and Marco Pavone$^{1}$% <-this % stops a space
\thanks{$^{1}$
Stanford University, USA {\tt\small\{aaryan04, gammelli, jthluke, gkarthik, pavone\}@stanford.edu}}%
\thanks{$^{2}$ Work done while at
ETH Zurich, Switzerland {\tt\small hedomini@student.ethz.ch}}
\iftoggle{extended}{
    \thanks{$^\dagger$ The project’s website can be found at: \url{https://github.com/StanfordASL/graph-rl-for-eamod}}
}{
}
}

\begin{document}

\maketitle
\thispagestyle{empty}
\pagestyle{empty}

%%%%%%%%%%%%%%%%%%%%%%%%%%%%%%%%%%%%%%%%%%%%%%%%%%%%%%%%%%%%%%%%%%%%%%%%%%%%%%%%

\begin{abstract}

Operators of Electric Autonomous Mobility-on-Demand (E-AMoD) fleets need to make several real-time decisions such as matching available vehicles to ride requests, rebalancing idle vehicles to areas of high demand, and charging vehicles to ensure sufficient range.
%Typically, the objective of the fleet operator is to maximize profits, which is defined as the revenue from serving rides minus the costs of operating the vehicles and charging them.
While this problem can be posed as a linear program that optimizes flows over a space-charge-time graph, the size of the resulting optimization problem does not allow for real-time implementation in realistic settings.
%In this work, we propose a graph reinforcement learning framework to improve the computational tractability of the E-AMoD control problem.
In this work, we present the E-AMoD control problem through the lens of reinforcement learning and propose a graph network-based framework to achieve drastically improved scalability and superior performance over 
\iftoggle{extended}{
    heuristics$^\dagger$.
}{
    heuristics.
    % $^\dagger$.
}
Specifically, we adopt a bi-level formulation where we (1) leverage a graph network-based RL agent to specify a \textit{desired next state} in the space-charge graph, and (2) solve more tractable linear programs to best achieve the desired state while ensuring feasibility.
Experiments using real-world data from San Francisco and New York City show that our approach achieves up to 89\% of the profits of the theoretically-optimal solution while achieving more than a 100x speedup in computational time.
We further highlight promising zero-shot transfer capabilities of our learned policy on tasks such as inter-city generalization and service area expansion, thus showing the utility, scalability, and flexibility of our framework.
Finally, our approach outperforms the best domain-specific heuristics with comparable runtimes, with an increase in profits by up to 3.2x.

\end{abstract}

%%%%%%%%%%%%%%%%%%%%%%%%%%%%%%%%%%%%%%%%%%%%%%%%%%%%%%%%%%%%%%%%%%%%%%%%%%%%%%%%
\section{INTRODUCTION}
Electric Autonomous Mobility-on-Demand (E-AMoD) systems use electric autonomous vehicles to provide on-demand ride-hailing services for customers. Operating an E-AMoD fleet involves three operations: matching available vehicles to customers who request rides, rebalancing idle vehicles to regions with high demand, and assigning vehicles to charging stations. 
In realistic settings, E-AMoD fleets can be centrally controlled and the operator can coordinate the assignment of vehicles to each of these three tasks to maximize efficiency, demand satisfaction, and profits.

%These systems can provide improved safety and convenience for customers, reduce congestion in urban areas by facilitating ride-sharing and be powered by sustainable energy resources \cite{metz2018developing, nunes2021estimating}. Furthermore, the fact that an entire fleet can be controlled by a single operator offers significant potential to coordinate all operations and improve the overall efficiency of the system. 

%In this paper, we consider the problem of profit maximization for an E-AMoD fleet operator. In particular, we seek to optimize three decisions: matching, rebalancing (or repositioning), and charging. First, the operator needs to match a set of ride requests to available vehicles while choosing rides that maximize profits and ensure that vehicles have sufficient range to complete their assigned rides. Next, if there are idle vehicles that are not serving customers currently, the operator has to also decide if and where they need to be rebalanced so that they may serve future demand effectively. Finally, the operator needs to determine which vehicles need to be charged, which charging stations they should go to, and for how long they should charge. 

It is worth emphasizing that these decisions need to be made in real time, and any offline schedule, even if computed at the start of a day using historical or predicted data, will generally be suboptimal \cite{turan2020dynamic} due to forecasting errors in traffic demand, vehicle energy consumption, and road congestion \cite{wen2019value}. Thus, we seek an approach to solve the E-AMoD control problem in real-time, so that we may be able to compute updated fleet coordination decisions whenever new information is available to the operator.

Among other approaches, Model Predictive Control (MPC) provides a framework to make decisions in a receding horizon fashion by repeatedly solving an optimization problem based on (i) the current state of the system and (ii) a forecast of future state elements. 
In the context of autonomous mobility-on-demand (AMoD) systems, receding-horizon control has been used extensively to make optimal rebalancing decisions \cite{zardini2022analysis}.
Specifically, prior work has shown that a network flow model for this problem can be successfully scaled for real-time large-scale operations \cite{tsao2018stochastic}. 
However, this approach assumes that all vehicles are \emph{indistinguishable} from each other, thereby enabling the operator to aggregate all vehicles and model their movements as a flow on a network. 
Electric vehicles on the other hand are \emph{distinguishable} based on their current state of charge, which determines their maximum range. 
Motivated by this, Estandia et al. \cite{EstandiaSchifferEtAl2019} use an augmented network flow formulation for E-AMoD systems, by including a charge dimension in the resulting graph (i.e., from a space-time graph to a space-\textit{charge}-time graph). 
However, the computational complexity of the resultant optimization problem does not allow for real-time implementation, and previous work has struggled to devise effective real-time controllers even for coarse representations of space, charge, and time. 
For instance, Estandia et al. \cite{EstandiaSchifferEtAl2019} report that solving the optimal control problem takes 42 minutes for an E-AMoD system operating across Orange County, USA over an eight-hour horizon.
% , while Rossi et al. \cite{RossiIglesiasEtAl2018b} report a 48 minute run time for a problem with 20 time steps, 30 charge levels, and 25 locations.

%in practice solving even modest size problems with 20 spatial zones, 30 time steps, and 30 discrete charge levels takes over 8 hours in a computer cluster.

%Our goal: vehicle level plan for EV
In this paper, we propose a strategy to design real-time controllers for E-AMoD systems through reinforcement learning.
To do so, we present the E-AMoD control problem through the lens of graph reinforcement learning (graph-RL) \cite{GammelliHarrisonEtAl2023} and exploit the main strengths of graph neural networks, reinforcement learning, and classical operations research tools.
%This motivates our research on using learning-based approaches to develop real-time controllers for E-AMoD control and coordination. In particular, we use graph reinforcement learning (RL) to decompose the larger MPC optimization problem into a sequence of simpler, RL-guided network-flow problems that can be easily solved in real-time. 

% can also use \textbf{} instead of \subsection to save space
\subsection{Related works}
Existing literature on the coordination of E-AMoD systems heavily relies on solving large-scale optimization problems.
Specifically, prior works approach the problem of joint optimization of charging station siting \cite{luke2021joint}, joint optimization of power flows \cite{EstandiaSchifferEtAl2019}, and the computation of optimal rebalancing plans \cite{bang2022congestion}.
However, in practice, the adoption of these methods is typically limited by their computational complexity and is only considered in an offline setting, thus not immediately applicable within real-time operations.
To improve the scalability of control algorithms for E-AMoD systems, previous works adopt several learning-based techniques.
%RL approaches have also been used in E-AMoD settings outside the context of optimal rebalancing. 
For example, Wan et al. \cite{wan2018model} consider charge scheduling for personal electric vehicles.
Bogyrbayeva et al. \cite{bogyrbayeva2021reinforcement} optimize nightly rebalancing operations of electric vehicles to charging stations, while Shi et al. \cite{shi2019operating} propose a decentralized algorithm for the charging-constrained vehicle routing problem.
Overall, although the aforementioned works cover a wide range of algorithms, there lacks a framework to deal with the joint computation of both spatial rebalancing and charging decisions within large-scale E-AMoD systems, which is a key factor in making the optimization problem prohibitively expensive and a main focus of this work.

Reinforcement learning has also been extensively used to learn fleet coordination policies that do not account for charging (i.e., AMoD systems).
For example, Gueriau et al. \cite{gueriau2018samod} developed RL-based decentralized approaches where the action of each vehicle is determined independently through a Q-learning policy, while Holler et al. \cite{HollerVuorioEtAl2019} developed a cooperative multi-agent approach for order dispatching and vehicle rebalancing using Deep Q-Networks and Proximal Policy Optimization.
Of particular relevance to our work are methods that (i) leverage the graph structure of the underlying transportation system, and (ii) combine principled control strategies with learned components in a hierarchical way \cite{GammelliYangEtAl2021, yu2021deep, GammelliYangEtAl2022}.
%, as these have been shown to facilitate learning within online \cite{GammelliYangEtAl2021} and few-shot \cite{GammelliYangEtAl2022}learning scenarios.
In this work, we leverage the framework of graph-RL to include charging within the range of autonomous, real-time decisions, thus substantially increasing the set of application areas.

\subsection{Our contributions}
% This paper makes two main contributions.
\textbf{Contribution \#1.} We present the first reinforcement learning agent that jointly learns the charging and rebalancing decisions for an E-AMoD fleet. 
This is enabled by two key design choices.
First, we leverage the power of graph neural networks to capture both spatial and charge information across the system.
This is critical in devising RL agents that can propagate information between different regions of the transportation network before computing a centralized decision for the entire fleet. 
Second, we extend the framework presented by Gammelli et al. \cite{GammelliYangEtAl2021} to the E-AMoD setting and develop an approach that leverages the specific strengths of direct optimization and reinforcement learning through a hierarchical formulation, which is advantageous in learning a policy that is more effective, scalable, and generalizable.
%This allows us to formulate the E-AMoD planning problem as a decision-making problem on a graph. Next, we extend a framework presented in \cite{gammelli2021graph} to the E-AMoD setting which allows us to transform the edge-level decision-making problem into a node-level planning problem. This reduces the dimensionality of the action space of the RL agent from $O(n^2)$ to $O(n)$, where $n$ is the number of nodes on a graph, thereby increasing the scalability of training and reducing the run-time at inference.

\textbf{Contribution \#2.} We provide numerical experiments that demonstrate how our approach is highly performant, scalable, and robust to changes in operating conditions. 
In particular, through extensive comparisons with both classical optimization-based approaches and domain-specific heuristics, results highlight how our approach achieves close-to-optimal performance with drastic runtime improvements.
%the computational and profitability benefits of our approach over an MPC solution and domain-specific heuristics, respectively. In particular, experiments highlight how our approach achieves 88.8\% to 95.6\% profits as the optimal solution while achieving nearly 27X to 85X speedup during runtime, depending on case study and MPC horizon length. Our approach also outperforms real-time heuristics for this problem by 4\% to 18\%, indicating our approach's ability to smartly choose optimal fleet control actions as a function of the state observed in real-time. Importantly, our ablations depict how our approach is robust to the non-linear explosion of the computation time inherence to the MPC solutions.

\textbf{Contribution \#3.} This work highlights how policies learned through graph-RL exhibit a series of desirable properties of fundamental practical importance for any system operator. 
Specifically, results show interesting transfer performance of a trained agent in the context of (i) inter-city generalization (i.e., the agent is trained on one city and directly applied to another), and (ii) service area expansion (i.e., the agent is trained on a specific sub-graph and directly applied to additional areas of the city).
We further show how the transfer capabilities achieved by agents learned through graph-RL are crucial in enabling learning within large-scale instances. 

\section{BACKGROUND}\label{sec:backgorund}
In this section, we introduce key terminology and notation in the context of reinforcement learning (Section \ref{subsec:backgrounda}) and graph neural networks for network control (Section \ref{subsec:backgroundb}).

\subsection{Reinforcement Learning}\label{subsec:backgrounda}
%Reinforcement learning addresses the problem of learning to control a dynamical system from experience.
We refer to a Markov decision process (MDP) as a tuple $\mathcal{M} = \left(\mathcal{S}, \mathcal{A}, P, r, \gamma \right)$, where $\mathcal{S}$ is the state space, $\mathcal{A}$ is the action space, $P$ describes the dynamics of the system through the conditional probability distribution $P(\bs_{t+1} | \bs_t, \ba_t)$, $r : \mathcal{S} \times \mathcal{A} \rightarrow \mathbb{R}$ defines a reward function, and $\gamma \in (0,1]$ is a scalar discount factor.
From a reinforcement learning perspective, the final goal is to learn a policy defining a distribution over possible actions given states, $\pi(\ba_t | \bs_t)$
%i.e., in the E-AMoD case, a distribution over the desired number of idle vehicles given the current state of the fleet.
by maximizing the expected cumulative reward $J(\pi) = \mathbb{E}_{\tau \sim p_{\pi}(\tau)} \left[\sum_{t=0}^{H} \gamma^t r(\bs_t, \ba_t) \right]$, where the expectation is computed under the distribution over trajectories $p(\tau)$ induced by policy $\pi(\ba_t | \bs_t)$ and system dynamics $P(\bs_{t+1} | \bs_t, \ba_t)$.

\subsection{Graph Neural Networks for Network Control}\label{subsec:backgroundb}
Given a graph $\mathcal{G} = (\mathcal{V}, \mathcal{E})$, where $\mathcal{V} = \{v_i\}_{i=1:N_v}$ and $\mathcal{E} = \{e_k\}_{k=1:N_e}$ respectively define the sets of nodes and edges of $\mathcal{G}$, most current graph neural network models can be seen as methods attempting to learn a function taking as input (i) a $D$-dimensional feature description $\mathbf{x}_i$ for every node $i$ (typically summarized in a $N_v \times D$ feature matrix $\mathbf{X}$), (ii) a representative description of the graph structure in matrix form $\bA$ (typically in the form of an adjacency matrix), and produce an updated representation $\mathbf{x'}_i$ for all nodes in the graph.
As observed in \cite{GammelliHarrisonEtAl2023}, graph neural networks (GNN) represent an extremely advantageous choice within network optimization problems, for three main reasons. 
First, GNNs are permutation invariant operators\footnote{We will refer to a computation as permutation invariant if its output is independent of the ordering of its inputs.}. 
This is particularly relevant in the context of graphs, where nodes do not have a natural ordering and where non-permutation invariant computations would consider each ordering as fundamentally different, and thus have been shown to require an exponential number of input/output training examples before being able to generalize.
Second, GNNs are local operators.
This enables the same neural network architecture to be applied to graphs of different sizes.
Third, GNNs align with the type of computations required within network optimization problems, which has been shown to lead to better performance and increased data efficiency.
\begin{comment}
Particularly relevant for this work is the Graph Convolution Network (GCN) \cite{kipf2016semi}.
At its core, a graph convolutional operator describes a parametric function $f(\mathbf{X}, \bA)$ for efficient information propagation on graphs.
Specifically, a GCN defines the following propagation rule:
\begin{equation}
    \mathbf{X'} = f(\mathbf{X}, \bA) = \sigma\left(\hat{\bD}^{-\frac{1}{2}} \hat{\bA} \hat{\bD}^{-\frac{1}{2}} \mathbf{X} \bW\right), 
    \label{eq:gcn}
\end{equation}
with $\hat{\bA} = \bA + \bI$, where $\bI$ is the identity matrix, $\hat{\bD}$ is the diagonal node degree matrix of $\hat{\bA}$, $\sigma(\cdot)$ is a non-linear activation function (e.g., ReLU) and  $\bW$ is a matrix of learnable parameters.
Intuitively, graph convolutions describe a permutation-invariant propagation rule that updates node-level features by aggregating information from neighboring nodes through the parameterization of a shared local filter across all locations in the graph.
\end{comment}

\section{THE E-AMOD CONTROL PROBLEM}\label{sec:setup}

% \subsection{Notations:}

In this section, we introduce the charge-expanded network flow model characterizing the E-AMoD control problem. 
Towards this aim, we partition the region of operation for an E-AMoD fleet (e.g., a city) into a set of discrete non-overlapping regions denoted $\mathcal{A}$.
The time horizon is discretized into a set of discrete intervals $\mathcal{T}=\{1,2,\cdots, T\}$ of a given length $\Delta T$. 
We consider a set of equally spaced discrete charge levels for each vehicle denoted by $\mathcal{C} = \{1,...,C\}$, where $C$ is the highest charge level. 
When a vehicle travels from region $i$ to region $j$, it takes $l_{ij}$ time steps and loses $\eta_{ij}$ units of charge. 
While charging, a vehicle moves up $t_c$ discrete charge levels per time step. 
We assume that our E-AMoD fleet has a fixed set of $N$ autonomous electric vehicles and that every region $a\in\mathcal{A}$ has a charging station with a finite number of charging plugs, denoted as $N_c^a$, with $N_c^a>0$, which is reasonable for most E-AMoD settings. 
We denote customer demand from region $i\in\mathcal{A}$ to $j\in\mathcal{A}$ at time $t \in \mathcal{T}$ as $d_{ij}^{t}$ and define the arrival process of passengers for each origin-destination pair as a time-dependent Poisson process that is independent of the arrival processes of other origin-destination pairs and the coordination of E-AMoD fleets\cite{Daganzo1978}.
%The proposed approach, nevertheless, can be readily extended to consider other types of arrival processes.
We denote the total demand from region $i\in\mathcal{A}$  at time $t$ as $d_{i}^{t} = \sum_{j\in\mathcal{A}}d_{ij}^{t}$. 
Note that customers might request a ride within the same spatial region, i.e., $d_{ii}^{t}$ can be non-zero. 
%At every time step $t$, customers who request a ride may or may not be assigned a vehicle. 
We assume that a customer that 
%hence a request from region $a$ at time $t$ can only be served by an idle vehicle available in region $a$ at that time. 
is not assigned a ride within one time step will leave the system.
% (which results in lost revenue). 

We denote the cost for increasing the charge level of a vehicle by one discrete level at time $t$ to be $p_{e}^{t}$, assumed to be known.
% The time dependence captures the time-varying electricity prices which can be higher during peak hours. Retail electricity prices are typically deterministic and known apriori, which is what we assume in this model.
We denote the cost of operating a vehicle from region $i\in\mathcal{A}$ to region $j\in\mathcal{A}$ as $o_{ij}^{t}$.
This cost is a function of the distance traveled, captures the amortized cost of maintenance, and is assumed to be given. 
Finally, the revenue for the operator generated by serving a passenger traveling from region $i$ to $j$ at time $t$ is denoted by $\rho_{ij}^{t}$.
\subsection{The Space-Charge Graph}
\label{subsec:2a}
\begin{figure*}[t]
    \centering
    \includegraphics[width=0.85\linewidth]{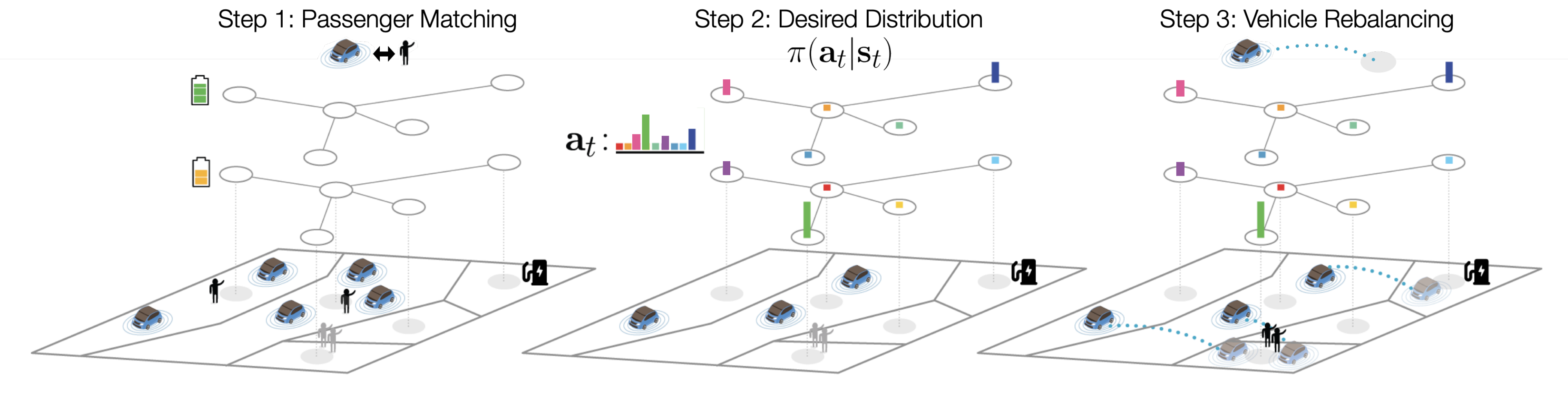}
    \caption{A visual representation of the tri-level framework for a given time step $t$. Step 1 (left) involves matching ride requests to vehicles. Step 2 (center) uses the policy learned through RL to compute an ideal redistribution of vehicles over the space-charge graph $\mathcal{G}$, i.e., $\mathbf{a}_{t}$. Step 3 (right) computes the spatial rebalancing and charging flows to achieve (as best as possible) the target distribution given by $\mathbf{a}_{t}$.}
    \label{fig:tri-level}
\end{figure*}

Having formally defined the relevant notation and assumptions used in this work, this section describes the graph structure characterizing the E-AMoD network flow problem (Figure \ref{fig:tri-level}).
%The E-AMoD control problem is naturally posed as a network flow problem, and we begin by describing the underlying graph structure of the problem. 
In this graph, nodes represent a \emph{(spatial region, charge level)} tuple and are used to capture the state of a vehicle. 
Multiple vehicles may have the same spatial region and charge level.
Over time, vehicles transition from one state to another as they satisfy passenger demands, perform spatial rebalancing, or get recharged. 
At any time, vehicles can transition from one node (i.e., spatial region and charge state) to another through edges that capture valid transitions. 
Formally, we denote the graph as $\mathcal{G} = (\mathcal{V}, \mathcal{E})$ where $\mathcal{V} = \mathcal{A} \times \mathcal{C}$ is the set of nodes and $\mathcal{E} = \{(i,j) \subset  \mathcal{V} \times \mathcal{V}\}$ is the set of edges. 
We denote an edge from node $i$ to node $j$ as $e_{ij}$. 
The graph $\mathcal{G}$ is characterized by two types of edges -- those representing the physical movement of vehicles from one region to another, denoted by $\mathcal{E}_{road}$, and those representing the charging of vehicles, denoted by $\mathcal{E}_{charge}$. Thus $\mathcal{E}=\mathcal{E}_{charge}\cup \mathcal{E}_{road}$. 
For any node $i \in \mathcal{V}$, we denote the corresponding spatial region and charge level as $i_r$ and $i_c$, respectively. 
Using this notation, we define $\mathcal{E}_{charge} = \{e_{ij}| i_r=j_r \text{ and } (j_c - i_c) = k t_{c} \text{ for some } k\in\mathbb{N} \}$. 
This set represents all possible transitions that can happen as a result of charging. 
Similarly, we define the edges corresponding to roads as $\mathcal{E}_{road} = \{e_{ij}| j_c=i_c- \eta_{ij} \}$. 
% We remark that this set can also include edges between nodes corresponding to the same spatial region with customers requesting rides within the region itself.
For every edge $e=(i,j) \in \mathcal{E}$ on the graph, we associate a time-varying cost $c_{ij}^{t}$ to traverse it. 
For road edges, $c_{ij}^{t} = o_{i_rj_r}^{t}$, while for a charging edges $c_{ij}^{t} = (j_c - i_c)p_{e}^{t}$. 
% Note that costs $c_{ij}$ are a function of $t$, as they depend on electricity price. 
% However, in what follows, we drop the explicit dependence on $t$ in the notation for simplicity.
Lastly, we define the travel time for all road and charging edges as $\tau_{ij}$ as $\tau_{ij} = l_{i_r j_r}$ and $\tau_{ij} = \ceil{\frac{(j_c - i_c)}{t_c}}$, respectively.

\subsection{The Optimal Control Problem}
The E-AMoD control problem is naturally posed as a network flow problem.
Formally, let $x_{ij}^{t}$ be the number of passengers who started traveling from $i\in\mathcal{V}$ to $j\in\mathcal{V}$ at time $t\in\mathcal{T}$, and let $y_{ij}^{t}$ be the number of vehicles that started rebalancing or charging at time $t$ from node $i$ to node $j$. 
The objective of the E-AMoD control problem is to maximize the profits over a pre-specified time horizon $T$, and is defined as follows:  
\begin{subequations}
\begin{align}
    & \max \quad  \sum_{t=1}^{T} \sum_{(i,j): e_{ij} \in \mathcal{E}} [ ( \rho_{ij}^{t} - c_{ij}^{t})x_{ij}^{t} - c_{ij}^{t} y_{ij}^{t}] \label{eq:global_obj} \\
    & \mathrm{s.\ t.} \quad   \nonumber \\
    & \sum_{(i,j)\in \mathcal{E}_{road}: i_r = u, j_r=v}x_{ij}^{t} \leq d_{uv}^{t} \quad \forall i,j\in\mathcal{V},  t\in\mathcal{T} \label{eq:global_c1}\\
    & \sum_{i\in\mathcal{V}}  (x_{ij}^{t - \tau_{ij}} + y_{ij}^{t - \tau_{ij}})  = \sum_{ k\in\mathcal{V}} (x_{jk}^{t} + y_{jk}^{t})  \; \forall t\in\mathcal{T}, j \in \mathcal{V} \label{eq:global_c2}\\
    & \sum_{(i,j):i_r = j_r = a} \sum_{k = 0}^{\tau_{ij}} y_{ij}^{t - k} \leq N_{c}^{a}  \quad \forall a \in \mathcal{A}, t\in\mathcal{T} \label{eq:global_c3}\\
    & x_{ii}^{0} = x^{\text{init}}_{i} \;,\; y_{ii}^{0} = 0 \quad \forall i\in\mathcal{V} \label{eq:global_c4}\\
    & x_{ij}^{t} \geq 0 \;,\; y_{ij}^{t} \geq 0, \label{eq:global_c5}
\end{align}
\end{subequations}
where, the objective term \eqref{eq:global_obj} represents the total profit, constraint \eqref{eq:global_c1} ensures that passenger flow does not exceed demand, \eqref{eq:global_c2} enforces flow conservation, \eqref{eq:global_c3} ensures that the number of vehicles charging at any point in time does not exceed the capacity of that station, and \eqref{eq:global_c4} and \eqref{eq:global_c5} set the initial conditions and ensure non-negativity of the decision variables, respectively. 

Note that the optimization problem in \eqref{eq:global_obj}-\eqref{eq:global_c5} involves $|\mathcal{E}|\times T$ decision variables. Since, the number of edges $|\mathcal{E}| = O(|\mathcal{V}|^2)$, and $\mathcal{V} = \mathcal{A} \times \mathcal{C}$, the number of decision variables in the problem is $O(|\mathcal{A}|^2 |\mathcal{C}|^2 T)$. Crucially, the rapid growth of the underlying optimization problem with respect to the spatial resolution, charge levels, and planning horizon $T$ results in poor computation performance for real-time applications, as observed in\cite{EstandiaSchifferEtAl2019}. 

%\subsection{Our Objective}
Our goal is to reduce the complexity of problem \eqref{eq:global_obj}-\eqref{eq:global_c5} to enable real-time control. 
To achieve this, we formulate the problem in \eqref{eq:global_obj}-\eqref{eq:global_c5} as a sequential decision-making problem \cite{fluri2019learning, GammelliYangEtAl2021}.
%,where we solve for $x_{ij}^{t}$ and $y_{ij}^{t}$ just for the current time $t$, and repeat this process at every time step \cite{fluri2019learning, gueriau2020shared, gammelli2021graph}. 
Our hypothesis is that we can express effective E-AMoD policies as a composition of policies: a higher-level policy trained through RL to maximize long-term reward, and a lower-level local approximation of the problem \eqref{eq:global_obj}-\eqref{eq:global_c5} to compute feasible, fleet-wide decisions.
This local correspondence is central to our formulation: we exploit exact optimization when it is useful, and otherwise push the complexities of optimizing for long-term performance to the learned policy.
%encode the temporal complexities of our current decisions to guide a computationally simpler local approximation of the problem \eqref{eq:global_obj}-\eqref{eq:global_c5}, we will not only be able to reduce the  computational cost, but also not compromise on the total profit. 
In the next section, we will present our approach via a tri-level framework, and discuss specific details about the design of our RL agent.

% \begin{table}[]
%     \centering
%     \begin{tabular}{c|c|c}
%         Variable & Description & Property \\
%          & & 
%     \end{tabular}
%     \caption{Caption}
%     \label{tab:my_label}
% \end{table}
\section{GRAPH-RL FOR E-AMoD CONTROL}\label{sec:method}

In this section, we introduce the proposed graph-RL framework for E-AMoD systems.
Specifically, we first describe a tri-level formulation used to approximate the problem \eqref{eq:global_obj}-\eqref{eq:global_c5} (Section \ref{subsec:3a}) together with the definition of a Markov Decision Process (MDP) within this formulation (Section \ref{subsec:3b}).
Lastly, we describe the proposed graph-RL agent (Section \ref{subsec:3c}).

\subsection{The Tri-level Framework} \label{subsec:3a}

Similar to \cite{fluri2019learning} and \cite{GammelliYangEtAl2021}, we approximate the problem in \eqref{eq:global_obj}-\eqref{eq:global_c5} through the following three-step decomposition: (i) derive passenger flows $x_{ij}^{t}$ by solving a matching problem, (ii) compute a desired distribution of idle vehicles across spatial and charge dimensions through reinforcement learning, and (iii) solve a minimal cost rebalancing problem to compute the desired flows $y_{ij}^{t}$ that would better achieve the desired distribution from the previous step (Figure \ref{fig:tri-level}).
Notice that solving this approximation drastically reduces the number of decision variables in the optimization problem.
Specifically, rather than solving a single planning problem with $|\mathcal{E}|\times T$ decision variables, we now solve three simpler problems that do not scale with the length of the planning horizon $T$, i.e., characterized by $|\mathcal{E}|$, $|\mathcal{V}|$, and $|\mathcal{E}|$  variables, respectively.
%This sequential decision-making framework, with a tri-level solution at each step drastically reduces the number of decisions that we need to take at each time step. 
%Thus, instead of solving one optimization problem with $|\mathcal{E}|\times T$ decision variables, we now solve three simpler problems that do not scale with the length of the planning horizon $T$. 
%The matching problem has $|\mathcal{E}|$ decision variables, the RL agent learns an action over all the $|\mathcal{V}|$ nodes, and the rebalancing solver optimizes over $|\mathcal{E}|$ variables to achieve the desired distribution. 
Crucially, we propose RL as an appealing learning paradigm to compensate for the lack of explicit planning caused by the three-step approximation and learn a control policy that optimizes long-term reward; in particular, we extend the framework proposed in \cite{GammelliYangEtAl2021} to handle \textit{space-charge} graphs (as opposed to \textit{space-only }graphs), developing RL-based approaches for hierarchical, large-scale network topologies.
%Furthermore, this formulation reduces the dimensionality of the action space for the RL agent to $O(|\mathcal{V}|)$, as it makes decisions on the ideal distribution of the vehicles, in contrast to an implementation where the RL agent learns the flows $x_{ij}^{t}$ and $y_{ij}^{t}$ and has an action dimension of $O(|\mathcal{E}|)$. The smaller dimension of the action space makes the learning problem easier for the RL agent, thereby improving performance. 
In what follows, we introduce the three-step framework in more detail.

\begin{comment}
We extend the three-step framework introduced in \cite{fluri2019learning, gueriau2020shared, gammelli2021graph} to control the E-AMoD fleets (see Fig. \ref{fig:tri-level}). At every time step, we first derive the passenger flows $x_{ij}^{t}$ by solving a matching problem. Second, we compute a desired spatial and charge level distribution for all idle vehicles after the matching step. For this step, we use a learned policy to compute the desired distribution over all nodes of the graph $G$. Finally, in the third step, we solve a minimal cost rebalancing problem to compute the desired flows $y_{ij}^{t}$ that would realize the desired distribution. In this framework, we have reduced the dimensionality of the action space of the problem to $|\mathcal{V}|$ (in contrast to an MPC solution which would require $O(|\mathcal{V}|^2|T)$ in edge-based approaches), increasing the scalability of the training and the run-time of the implementation. We will now explain the three-step framework to make a decision at every time step $t$ in more detail.
\end{comment}

\textbf{Step 1: Passenger Matching.}
The first step is passenger matching, wherein the following matching problem is solved to derive passenger flows:

\begin{subequations}
\begin{align}
    \max_{x_{ij}^{t} \geq 0} \quad & \sum_{(i,j): e_{ij} \in \mathcal{E}_{road}}(\rho_{ij}^t-c_{ij}^{t}) x_{ij}^t \label{eq:match_obj}\\
\mathrm{s.\ t.} \quad & \sum_{j : e_{ij} \in \mathcal{E}_{road}}{x_{ij}^t} \leq n_{init}^{t}[i] \quad \forall i \in \mathcal{V} \label{eq:match_c1}\\
    & \sum_{\substack{(i,j): e_{ij} \in \mathcal{E}_{road} \\ i_r =a_1, j_r = a_2}}{x_{ij}^t} \leq d_{a_1a_2}^t \quad \forall a_1, a_2 \in \mathcal{A}, \label{eq:match_c2}
\end{align}
\end{subequations}
where the objective term \eqref{eq:match_obj} denotes the difference between the revenue and cost of traversing all the edges, constraint \eqref{eq:match_c1} limits the maximum flow from each node to the number of available idle vehicles in node $i$ at time $t$ before matching, denoted as $n_{init}^{t}[i]$, and \eqref{eq:match_c2} ensures that the passenger flow between any two nodes does not exceed demand.
Notice that since the constraint matrix is totally unimodular, the resulting passenger flows are positive integers, i.e., $x_{ij}^t\in \mathbb{Z}_+$ if $d_{ij}^t \in \mathbb{Z}_+,~\forall i,j\in\mathcal{V}$.
%Also note that a trade-off of this decomposition is that passengers are matched to vehicles without consideration of the vehicles' charge level.
%Note that the constraint matrix for this optimization problem is totally unimodular \cite{gammelli2021graph}, and the resultant passenger flows $x_{ij}^{t}$ will all be integral if the demand $d_{a1, a2}^{t}$ are integral. %Also, replacing this matching step with a learning-guided approach to improve performance is a direction of future research.

\textbf{Step 2: Desired Distribution.}
The second step entails determining the desired number of idle vehicles $n_{target}^{t}[i]$ at each node.
Let us denote the number of idle vehicles at node $i\in\mathcal{V}$, after the matching step as $n_{idle}^{t}[i]$, i.e., $n_{idle}^{t}[i] = n_{init}^{t}[i] - \sum_{j\in\mathcal{V}}x_{ij}^{t}$.
%These are the number idle at time $t$ vehicles for which we need to make a rebalancing or charging decision. 
%Note that $n_{idle}^{t}[i] = n_{init}^{t}[i] - \sum_{j\in\mathcal{V}}x_{ij}^{t}$.
In this work, we compute $n_{target}^{t}[i]$ in two steps.
First, we determine a desired \textit{distribution} of vehicles (i.e., the action for the RL agent) $\mathbf{a}_{t} = \{a_{t}[i]\}_{i \in \mathcal{V}}$, where $a_{t}[i] \in [0,1]$ defines the percentage of currently available vehicles to be moved to node $i$ at time $t$, and $\sum_{i \in \mathcal{V}} a_{t}[i] = 1$. 
Second, we use the desired distribution to compute $n_{target}^{t}[i] = \left\lfloor a_{t}[i] \sum_{i \in \mathcal{V}}n_{idle}^{t}[i] \right\rfloor$ as the actual number of vehicles at each node $i$.
%In the second step, we use a learning agent to identify an action vector $\mathbf{a}_{t} \in \mathbb{R}^{|V|\times 1}$ over all the nodes of the graph $G$. 
%The action vector $\mathbf{a}_{t}$ is a probability density function for the desired distribution of idle vehicles.
%Thus $\sum_{i\in\mathcal{V}}a_{t}[i] = 1$, and the desired number of vehicles at any node $i$, $n_{target}^{t}[i] = \left\lfloor a_{t}[i] \sum_{i \in \mathcal{V}}n_{idle}^{t}[i]  \right\rfloor$. 
Here, the floor function $\lfloor\cdot\rfloor$ ensures that the desired number of vehicles is integer and always available.
It is important to highlight how this action representation is scale-invariant by construction, as it acts on \textit{ratios} as opposed to raw vehicle counts: a strategy that has been shown to lead to increased learning stability and better generalization \cite{GammelliYangEtAl2021}.
Crucially, the goal of our formulation is to use reinforcement learning to learn a policy over desired distributions $\mathbf{a}_{t}$ that is capable of steering the myopic behavior of steps 1 and 3 towards long-term optimality.
%Observe that our agent learns the distribution of vehicles, rather than the actual number of vehicles. 
%This is because prior works \cite{gammelli2021graph} have shown that the use scale-invariant action representations, opposed to raw commodity quantities, learning is easier and the learned model is more generalizable. %We will discuss further details about the learning agent in the next subsection. 

\textbf{Step 3: Vehicle Rebalancing.} 
Henceforth, we refer to ``rebalancing'' as both spatial rebalancing and charge rebalancing (i.e., charging) for brevity. In this third step, we use a linear program to compute the rebalancing flows $y_{ij}^{t}$ that (i) achieve the desired distribution from step 2 in the minimum cost, and (ii) satisfy domain-specific constraints. 
Specifically, this is achieved by solving the following problem:
%While we wish to rebalance vehicles exactly as suggested by the RL agent to maximize rewards, it may not always be possible. In particular, since the graph is not fully connected, we may not be able to move vehicles to arbitrary states by just taking a decision to move through one edge at time $t$. For example, we cannot rebalance a vehicle to lose all its charge in one time step, irrespective of the RL agents action. Thus, this optimization problem is designed to use the RL agents action as a guide while ensuring that any poor decision by the learned agent does not lead to infeasible decisions. For this formulation, we introduce a non-negative slack variable $s_v \forall v \in \mathcal{V}$ to capture how closely we can match the target accumulation for each node, and use it to penalize deviations from the target distribution while ensuring feasibility. The resultant formulation is presented in Eqns. \eqref{eq:rebalance_obj}-\eqref{eq:rebalance_c2}.
\begin{subequations}
\begin{align}
\min_{ \substack{y_{ij}^{t} \geq 0 \\ s_v \geq 0 }} &  \quad \sum_{(i,j): e_{ij} \in \mathcal{E}} c_{ij}^{t} y_{ij}^t + M \sum_{v \in \mathcal{V}}s_{v} \label{eq:rebalance_obj}\\
\mathrm{s.\ t.} & \quad \sum_{j \in \mathcal{V}}{y_{ij}^t} \leq n_{idle}^{t}[i] \quad, \forall i \in \mathcal{V} \label{eq:rebalance_c1}\\
& \quad \sum_{i \in \mathcal{V}}(y_{ij}^t - y_{ji}^t) + s_j \notag \\
& \quad \quad = n_{target}^{t}[j]  - n_{idle}^{t}[j]  \quad, \forall j \in \mathcal{V}, \label{eq:rebalance_c2}
\end{align}
\end{subequations}
where the objective term \eqref{eq:rebalance_obj} represents the rebalancing cost plus a penalty for deviations from the desired distribution, with $s_v$ defined as a slack variable for vehicle deviation and $M$ as a large penalty factor, constraint \eqref{eq:rebalance_c1} limits the rebalancing flows from a region to the vehicles available in that region, and \eqref{eq:rebalance_c2} ensures that the resulting number of vehicles (the left-hand side) is close to the desired number of vehicles (the right-hand side).

\subsection{The E-AMoD Markov Decision Process} \label{subsec:3b}
%Reinforcement learning offers a technique to control a Markov decision process (MDP) by observing actions and their rewards. An MDP is defined by the tuple $(S, A, P, r)$, where $S$ is the set of states, $A$ is the set of actions, $P$ describes the dynamics of the system through a conditional probability distribution function, and $r: S\times A \rightarrow \mathbb{R}$ is the reward function. Thus, an action $\mathbf{a}_{t} \in A$ on the current state $\mathbf{s}_t \in S$ gives a reward $r(\mathbf{s}_t, \mathbf{a}_t)$ and the next state is described by the pdf $P(\mathbf{s}_{t+1}| \mathbf{s}_t, \mathbf{a}_t)$. The objective of an RL agent is to learn a policy $\pi_{\theta}:S\rightarrow A$ parameterized by $\theta$ to maximize the expected long term discounted reward $J(\theta) = \mathbb{E}_{\mathbf{a}_t \sim \pi_{\theta}(\mathbf{s}_t)}\left[  \sum_{t=1}^{T} \gamma^t  r(\mathbf{s}_t, \mathbf{a}_t) \right]$, where $\gamma \in (0,1)$ is the discount factor.
In this section, we formulate the E-AMoD control problem as an MDP.
%Specifically, we aim to learn a behavior policy to select the desired distribution introduced in Section \ref{subsec:3a} (Step 2).
In what follows, we define the elements characterizing the MDP for the E-AMoD control problem.

\textbf{State space.}
We define the state of the system $\mathbf{s}_t$ to capture relevant information required to express effective fleet control strategies.
To do so, we define the state representation to encode information about (i) the topology of the space-charge network through an adjacency matrix $\mathbf{A}$, and (ii) local information about each node in the network through a feature matrix $\mathbf{X}$.
On one hand, the topology of the space-charge network is fully characterized by the adjacency matrix $\mathbf{A}$ of graph $\mathcal{G}$, as introduced in Section \ref{subsec:2a}.
On the other, we choose the feature matrix $\mathbf{X}$ to be defined by three main sources of information. 
Firstly, we characterize the state of the E-AMoD system by the current and \textit{projected} number of idle vehicles across all nodes, $n_{idle}^{t}[i]$ for $t=t, \hdots, t = t+K, \forall i \in \mathcal{V}$.  
%Such information helps the agent account for vehicles that may become available in a certain area due to prior decisions.
Note that the projected number of idle vehicles is readily estimated given past matching and rebalancing actions (i.e., $x_{ij}^{k}$ and $y_{ij}^{k}$ for all $k<t$), as well as travel times $\tau_{ij}$.
Secondly, profit-maximizing control policies will necessarily depend on information regarding the potential revenue that can be obtained across different regions. 
To do so, we express the potential revenue across all regions $i$ over the next $K$ time steps as the sum of all revenues from estimated trips originating from that region: $\left(\sum_{j \in \mathcal{A}} \hat{d}_{ij}^{t+1}\hat{\rho}_{ij}^{t+1}, \hdots, \sum_{j \in \mathcal{A}} \hat{d}_{ij}^{t+K}\hat{\rho}_{ij}^{t+K}\right)$, where $\hat{d}_{ij}^{t+1}$ and $\hat{\rho}_{ij}^{t+1}$ are the estimated demand and trip revenue between regions $i$ and $j$. 
Finally, to enable proactive charging policies, we distinguish between nodes at different charge levels through the fractional charge level $\frac{i_c}{|\mathcal{C}|}$ of the node.

\textbf{Action space.}
In this work, we consider the problem of learning a \textit{desired distribution of idle vehicles} across all nodes in the graph $\mathbf{a}_t \in \mathbb{R}^{|\mathcal{V}|}_{\geq 0}$.
Specifically, we define the action $\mathbf{a}_t$ to describe a probability distribution over vehicle charge level and location.
%The action vector $\mathbf{a}_t \in \mathbb{R}^{|\mathcal{V}|\times 1}_{\geq 0}$ is a probability distribution over all nodes of the graph $G$. After rescaling $\mathbf{a}_{t}$ with the total number of idle vehicles, $\sum_{i\in\mathcal{V}}n_{idle}^{t}[i]$, the action represents the target number of vehicles that we should have at each spatial region and each charge level. Note that the RL agent should learn actions that result in rebalancing decisions that are beneficial in the long run, to compensate for the myopic sequential decision making that is introduced by our tri-level decomposition.

\textbf{Reward.}
We define the reward function for the MDP such that the RL agent learns actions that maximize the global objective described in \eqref{eq:global_obj}. 
To do so, the instantaneous reward should reflect the revenue from passenger trips as well as the rebalancing costs associated with fleet management.
Specifically, given trip revenues and costs (i.e., $\rho_{ij}^{t}$ and $c_{ij}^{t}$) together with the number of passenger and rebalancing trips (i.e., $x_{ij}^{t}$ and $y_{ij}^{t}$) we define the reward as:
\begin{equation}
    r(\mathbf{s}_t, \mathbf{a}_t) = \sum_{i,j} \left[ (\rho_{ij}^{t}-c_{ij}^{t}) x_{ij}^{t} - c_{ij}^{t} y_{ij}^t \right].
\end{equation}
%Note that although this reward only includes the profit at time $t$, the RL agent will learn to take actions that maximize the discounted cumulative reward over the entire horizon $T$. 

\textbf{Dynamics.}
The dynamics characterizing the E-AMoD MDP describe both the stochastic evolution of the system, as well as how fleet management decisions influence future state elements, such as the availability and distribution of idle vehicles.
Specifically, we assume the evolution of travel demand between regions $d_{ij}^t$ to be independent of the operator decisions and follows a time-dependent Poisson process (in our experiments, estimated from real trip travel data).
On the other hand, some of the state variable's transitions deterministically depend on the chosen action.
For example, the projected availability $n_{idle}^{t'}[i], \forall i \in \mathcal{V}, t' > t$ is uniquely defined as the sum of the current availability $n_{idle}^{t}[i]$ together with the projected number of incoming vehicles at time $t'$ (from both passenger and rebalancing trips), minus the vehicles currently chosen to be rebalanced.
Finally, state variables related to provider information, such as trip price $\rho_{ij}^t$ and cost $c_{ij}^{t}$ are assumed to be exogenous and known (hence, independent from the actions selected by the operator). 

%The transition probability function describes how the action affects the evolution of the state. Recall that we had three components to the state description at each node: the present number of idle vehicles, the future number of idle vehicles, and the projected revenue (i.e., demand multiplied. by the average price of the trip). The present and the future number of idle vehicles at a node depends on the action vector as $\mathbf{a}_t$ affects the subsequent rebalancing and matching decisions that directly impact idle vehicle distributions. The future revenue component of the state is however exogenously determined in our experiments and not dependent on the action. In our MDP, we draw the future demand from a fixed distribution (obtained using real data) and solve two optimization problems to draw a sample from this state transition distribution $P$ for a given $(\mathbf{s}_t, \mathbf{a}_t)$ pair.

\subsection{Graph-RL Agent} \label{subsec:3c}
After having introduced the E-AMoD control problem and the related MDP formulation, we now formally describe the graph network-based architecture characterizing the proposed RL agent.
In this work, we learn E-AMoD control policies through the Soft-Actor-Critic (SAC) \cite{HaarnojaEtAl2018} algorithm and define the following architectures for policy (i.e., the actor) and value function estimator (i.e., the critic).

\textbf{Actor.} 
As described in Section \ref{subsec:3b}, the goal of the policy network is to learn a mapping from the current state of the system $\bs_t$ to a desired distribution of idle vehicles $\ba_t$.
We define $\pi(\ba_t | \bs_t)$ as a Dirichlet distribution over nodes in the graph (i.e., $\ba_t \sim \pi(\ba_t | \bs_t) = \text{Dir}(\ba_t | \mathbf{\alpha}_t)$), with the policy network parametrizing the concentration parameters $\mathbf{\alpha}_t \in \mathbb{R}_{+}^{|\mathcal{V}|} $.
The neural network used in our implementation consists of one layer of graph convolution network \cite{kipf2016semi} with skip-connections and ReLU activations, whose output is then aggregated across neighboring nodes using a sum-pooling function, and finally passed to three MLP layers of 32 hidden units to produce the Dirichlet concentration parameters.

\textbf{Critic.}
The architecture of the critic mostly overlaps with the one used to define the policy network.
The main difference between the two architectures lies in an additional \emph{global} sum-pooling performed on the output of the graph convolution to compute a single value function estimate for the entire network, i.e., opposed to the actor that computes an output for every node. 

%After processing node and edge features via the same graph convolutional layer described above, the value function concatenates output with the original unprocessed node features and the action specified. Two subsequent MLP layers process the concatenated output, each passing its respective outputs through ReLU activation functions. The output of the second ReLU activation function is condensed via a sum-aggregation across all nodes before being passed into the final MLP layer. The architecture of the value function is also independent of the size of the graph, thereby allowing for training and testing over graphs of varying sizes.  

\section{Experiments}\label{sec:expt}
In this section, we present simulation results that demonstrate the performance of our proposed approach.
Specifically, the goal of our experiments is to answer the following questions: (1) Can the proposed graph-RL framework learn effective fleet management strategies in real-world urban mobility scenarios? (2) Computationally, what are the advantages of graph-RL approaches compared to traditional optimization-based strategies and domain-specific heuristics? (3) What are the generalization capabilities of behavior policies learned through our approach?

\subsection{Simulation Environment and Baselines}
We model an E-AMoD system serving passenger travel in San Francisco and New York City for 12 hours, from 8am-8pm.
In San Francisco, travel demand $d_{ij}^t$ and travel times $l_{ij}$ is based on origin-destination travel data for all passenger travel for an average weekday in 2019, provided by StreetLight Data\footnote{\url{https://www.streetlightdata.com/}}.
We use this data to calibrate a Poisson process of travel demand, which we use to generate unseen (but realistic) demand patterns for both training and test scenarios.
In New York City, travel demand and travel times are based on High Volume For-Hire Vehicles origin-destination data
%in MONTH of YEAR,
provided by the New York City Taxi and Limousine Commission\footnote{\url{https://www.nyc.gov/site/tlc/index.page}}.
Each spatial region is approximately 6 traffic analysis zones (TAZ).
In each experiment, we assume fleet size is 20\% of the peak total travel demand.
Additionally, we assume there are a total number of 50 kW charging stations equal to 20\% of the fleet size, which are distributed uniformly across all spatial regions to determine $N_c^a$.
We model the fleet vehicle based on the Chevrolet Bolt EV, with 65 kWh and an energy consumption of 0.4037 kWh/mi that includes a 30\% de-rating due to charging losses and autonomous vehicle auxiliary loads, which we use to determine $\eta_{ij}$.
Reserving 40\% of the vehicle battery capacity for operational uncertainty in energy consumption, and setting the charge level step size to be 2 kWh,
% , close to the first quartile of trip energy consumption,
we result in $C=19$ charge levels.
Setting $\Delta T=15$ minutes, we have $t_c=6$ charge levels.
The electricity price $p_e^t$ is based from Pacific Gas \& Electric's Business Electric Vehicle\footnote{\url{https://www.pge.com/tariffs/assets/pdf/tariffbook/ELEC_SCHEDS_BEV.pdf}} time-of-use energy rates in 2022, which promotes charging when solar generation is most plentiful.
Its rates are 0.16872 \$/kWh from 8am-9am and 2pm-4pm, 0.14545 \$/kWh from 9am-2pm, and peak price of 0.38195 \$/kWh from 4pm-8pm.
The amortized cost of maintenance $o_{i_r,j_r}$ is calculated using 0.077 \$/mi from the American Automobile Association\footnote{\url{https://newsroom.aaa.com/wp-content/uploads/2021/08/2021-YDC-Brochure-Live.pdf}}.
The revenue from serving passengers $\rho_{ij}^{t}$ is based on Lyft pricing for the San Francisco Bay Area\footnote{\url{https://www.lyft.com/pricing/SFO}} in 2022, with a base fare and service fee of \$4.90, price per mile of 0.90 \$/mi, and price per minute of 0.39 \$/min.

In our experiments, we compare the proposed graph-RL framework with heuristic and MPC-based methods. All three approaches are targeting to solve the E-AMoD control problem within a real-time constraint of 10 seconds \cite{luo2019dynamic}, which may be demanded by a lower-level vehicle dispatch algorithm. The heuristic and MPC-based methods are implemented as follows:

\noindent \textbf{Heuristics.} We focus on measuring the performance of realistic, domain-specific fleet management heuristics. This class of methods also adopts the tri-level framework outlined in \ref{subsec:3a}, but determines the desired distribution in Step 2 heuristically.

First, vehicles are recharged using one of the following methods:
\begin{enumerate}
        \item \emph{Empty-to-full}: all vehicles that reach a charge level below the average trip's energy consumption are recharged to full.
        \item \emph{Off-peak Absolute}: when electricity price is not at its peak, all vehicles below 30\% charge level are recharged for one time step (i.e., recharge $t_c$ charge levels). During peak price, vehicles with charge level below the average trip's energy consumption are recharged for one time step.
        \item \emph{Off-peak Relative}: equivalent to Off-peak Absolute except that during off-peaks the lowest 30\% of vehicles by charge level in each region are recharged.
\end{enumerate}
Second, idle vehicles are spatially rebalanced to uniformly distribute them across all spatial regions.
These benchmarks provide a measure of performance for methods that are computationally feasible and simplest to implement by real-world operators.

\noindent \textbf{MPC-based.} Within this class of methods, we focus on measuring performance of MPC approaches that serve as state-of-the-art benchmarks for the E-AMoD control problem. 

\begin{enumerate}
    \setcounter{enumi}{3}
    \item \emph{MPC-Oracle}: the MPC is based on problem \eqref{eq:global_obj}-\eqref{eq:global_c5} that assumes perfect foresight information of future user requests for all time steps.
    This approach serves as an \emph{oracle} that provides a performance upper bound for any fleet management policy. Notice that MPC-Oracle does not scale well as the number of regions increases, since solving the optimization model is extremely computationally expensive.
    \item \emph{MPC}: we relax the assumption of perfect foresight information in MPC-Oracle. Additionally, in an attempt to approach the real-time constraint of 10 seconds, the planning horizon is reduced to three time steps of look-ahead for scenarios with 5 spatial regions, and a single time step look-ahead for scenarios with 10, 15, and 20 spatial regions.
    This approach is a more realistic optimization-based benchmark, but might still violate real-time constraints.
\end{enumerate}
In our experiments, we monitor a set of metrics not directly included in the reward function.
Specifically, we report performance with respect to (i) \textit{Served demand}: defined as the number of completed passenger trips, (ii) \textit{Operating cost}: defined as the overall cost induced on the system by non-passenger trips (i.e., accounting for both charging and spatial rebalancing), and (iii) \textit{Percentage of oracle performance}.

\subsection{Learning to Control E-AMoD Fleets}
\begin{figure}
  \includegraphics[width=0.95\linewidth]{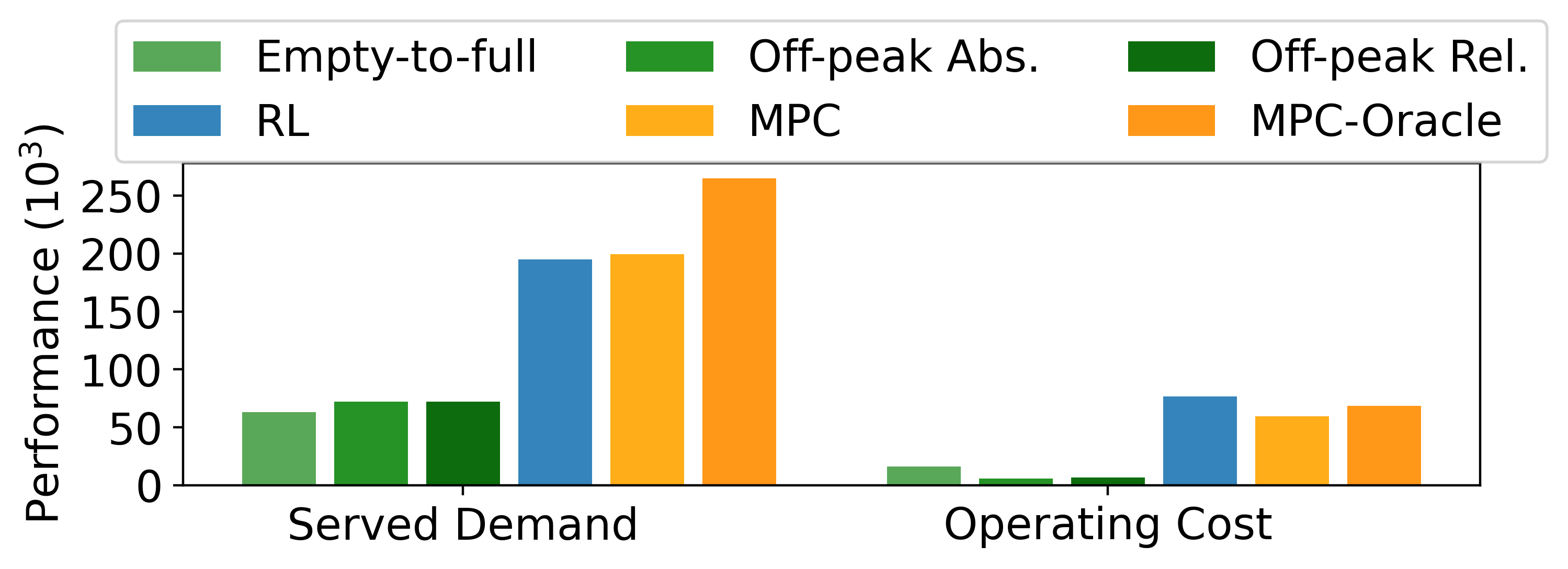}
\caption{Average served demand and operational cost comparison on San Francisco and New York (5, 10, 15, 20) scenarios.}
\label{fig:kpi}
\end{figure}

\begin{table*}[t]
%\small	
\centering
\begingroup
\renewcommand*{\arraystretch}{1.25}
\begin{tabular}{l l l c c c c c c}
 & & & \multicolumn{3}{c}{Heuristics} & RL & \multicolumn{2}{c}{Optimization} \\
 & & & Empty-to-full & Off-peak Abs. & Off-peak Relative & Graph-RL (ours) & MPC & MPC-Oracle\\
\hline
 \multirow{8}{1em}{A} & \multirow{4}{6em}{San Francisco} & 5 & 32.1\% (0.5s) & 32.1\% (0.5s) & 32.2\% (0.5s) & 87.4\% (0.4s) &  \textbf{\highest{94.2\% (9.7s)}} & 971.7  \,\,(1:41min)\\
 & & 10 & 27.0\% (1.6s) & 30.5\% (1.6s) & 30.6\% (1.6s) & \highest{81.6\% (1.9s)} & \textbf{88.4\% (15.2s)} & 2271.2 \,(6:51min)\\
 & & 15 & 27.9\% (4.4s) & 31.9\% (4.4s) & 32.0\% (4.4s) & \textbf{\highest{76.4\% (3.5s)}} & 67.7\% (33.3s) & 3544.8 (16:01min) \\
 & & 20 & 26.7\% (6.5s) & 30.7\% (6.6s) & 30.7\% (6.7s) & \textbf{\highest{78.1\% (7.8s)}} & 73.6\% (58.6s) & 5261.9 (28:47min)\\
& \multirow{4}{6em}{New York} & 5 & 26.6\% (0.4s) & 30.5\% (0.4s) & 30.5\% (0.4s) & 89.0\% (0.5s) & \textbf{\highest{95.1\% (9.6s)}} & 84.5 \,\,\,\,\,(1:37min)\\
 & & 10 & 18.8\% (1.8s) & 22.5\% (1.8s) & 22.9\% (1.8s) & \highest{74.3\% (1.8s)} & \textbf{81.5\% (14.4s)} & 292.1 \,\,\,(6:19min)\\
 & & 15 & 18.7\% (4.2s) & 22.3\% (4.2s) & 22.3\% (4.2s) & \highest{63.4\% (3.3s)} & \textbf{79.2\% (31.6s)} & 528.6 \,(14:34min)\\
 & & 20 & 18.1\% (7.5s) & 21.7\% (7.6s) & 21.7\% (7.7s) & \highest{55.5\% (7.39s)} & \textbf{75.0\% (54.4s)} & 930.8 \,(24:58min)\\
\hline
  \multirow{8}{1em}{B} & \multirow{4}{6em}{SF $\rightarrow$ NY} & 5 & - & - & - & 77.0\% (0.5s) & - & - \\
 & & 10 & - & - & - & 35.6\% (1.8s) & - & - \\
 & & 15 & - & - & - & 54.7\% (4.0s) & - & - \\
 & & 20 & - & - & - & 30.3\% (7.5s) & - & - \\
 & \multirow{4}{6em}{NY $\rightarrow$ SF} & 5 & - & - & - & 55.1\% (0.5s) & - & - \\
 & & 10 & - & - & - & 48.8\% (1.9s) & - & - \\
 & & 15 & - & - & - & 46.1\% (4.1s) & - & - \\
 & & 20 & - & - & - & 44.5\% (7.6s) & - & - \\
 \hline
  \multirow{3}{1em}{C} & \multirow{3}{6em}{San Francisco} & 5 $\rightarrow$ 20 & - & - & - & 47.7\% (7.7s) & - & - \\
 & & 10 $\rightarrow$ 20 & - & - & - & 66.7\% (7.6s) & - & - \\
 & & 15 $\rightarrow$ 20 & - & - & - & 74.6\% (7.7s) & - & - \\
 \hline
\end{tabular}
\caption{Percentage of oracle reward (profit, thousands of dollars) and computation time per decision on test scenarios. \textbf{Black-bold} and \highest{red} highlight the best-performing (non-oracle) model and the best-performing model that satisfies real-time constraints (i.e., 10 sec.), respectively. \textbf{\highest{Red-bold}} is used in case the two coincide.}
\label{tab:exp1}
\endgroup
\end{table*}
In our first simulation experiment, we study system performance on both San Francisco and New York scenarios, across increasing spatial coverage (i.e., from 5 spatial regions to 20).
Results in Table \ref{tab:exp1}, Part A show that policies learned through graph-RL achieve, on average, $\approx 75\%$ of MPC-Oracle, which assumes perfect foresight of future demand and unlimited computation time.
Table \ref{tab:exp1} and Figure \ref{fig:kpi} also highlight how the proposed approach is comparable in performance with the more realistic implementation of MPC, outperforming it in the SF15, SF20 scenarios and with a slight loss in performance on NY5, NY10 and SF5, SF10.
Crucially, despite it being outperformed by MPC in some of the above instances, the graph-RL policy is the only method (together with the three heuristics) that can successfully satisfy the computation time constraints: this is of critical importance for the operator, as it would simply not be able to execute MPC in real-time beyond 5 spatial regions.
Thus, results in Table \ref{tab:exp1}, Part A and Figure \ref{fig:kpi} show how graph-RL is able to maintain the performance of optimization-based methods, while substantially outperforming heuristics with comparable computation time. 

\smallskip
\noindent \textbf{Computational analysis.} 

\begin{figure}[t]
  \includegraphics[width=0.95\linewidth]{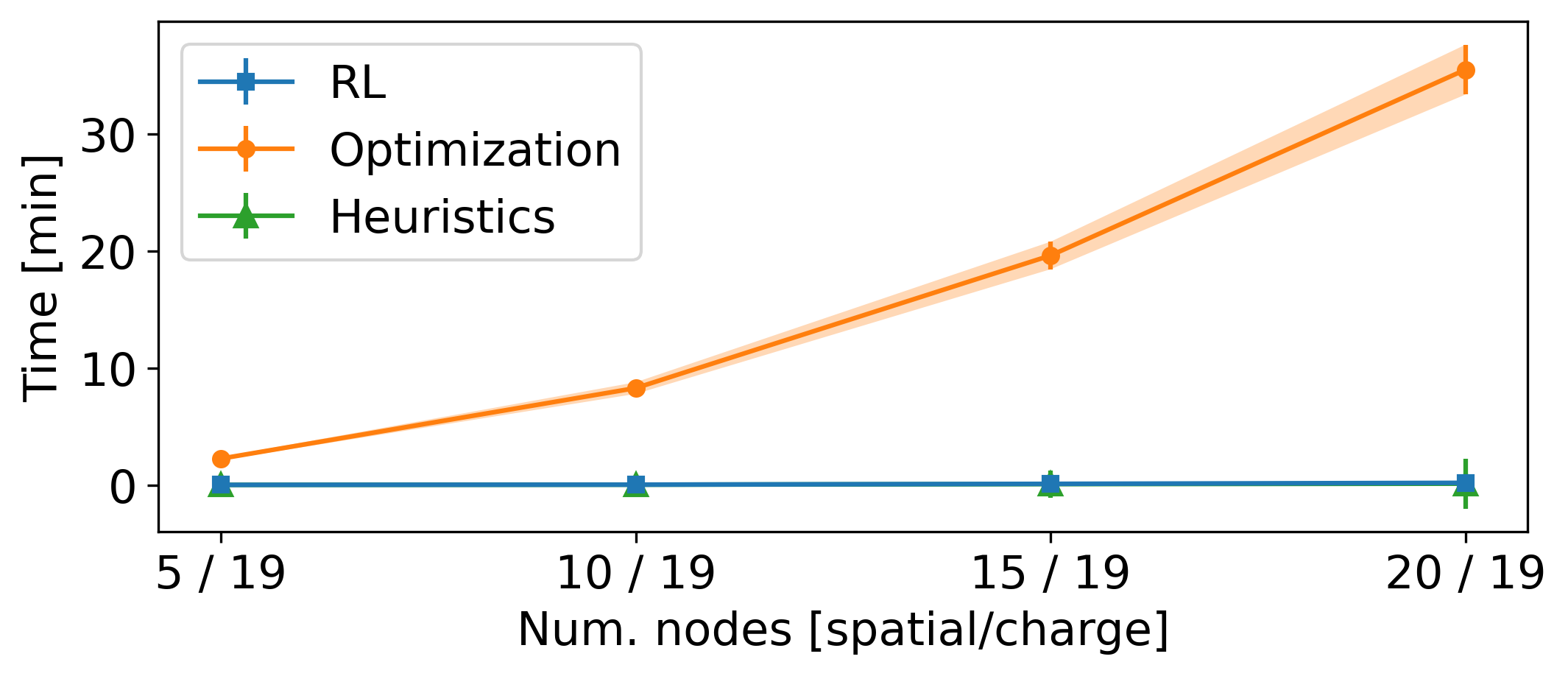}
\caption{Comparison of computation times between optimization (MPC-oracle, orange), graph-RL (blue), and heuristics (green).}
\label{fig:computational_analysis}
\end{figure}

After having discussed system performance in the previous experiment, we further study the computational cost of the proposed graph-RL approach compared to both heuristics and optimization-based approaches.
As shown in Fig \ref{fig:computational_analysis}, we compare the time necessary to compute a single decision across varying dimensions of the underlying space-charge graph, ranging from 5 spatial regions and 19 charge levels up to 20 spatial regions and 19 charge levels.
The results show how policies learned through graph-RL are approximately on par with heuristics, as opposed to optimization-based methods that scale super-linearly with the dimensionality of the problem.
In practice, we compare the proposed graph-RL approach to (i) Off-peak Relative as a representative heuristic, as all heuristics considered in this work have comparable runtime, and (ii) MPC-Oracle to highlight the theoretical complexity of the underlying control problem.
Crucially, Fig \ref{fig:computational_analysis} highlights the appealing scalability of RL-based methods and shows that learning-based approaches allow for real-time control by forward-propagation of the current system state through the learned policy $\pi(\ba | \bs)$, essentially amortizing the cost of optimization.

\subsection{Transfer and Generalization}
\smallskip
\noindent \textbf{Inter-city transfer.} 
To assess the transfer capabilities of graph-RL within E-AMoD systems, we also study the degree to which a policy learned on one city can be applied \textit{zero-shot} (i.e., without further training) to a different city.
Concretely, we do so by selecting a policy trained in New York and then deploying it in San Francisco (and vice-versa). 
As for the rest of our experiments, we repeat this procedure across varying levels of spatial coverage.
Despite the lower overall performance, results in Table \ref{tab:exp1}, Part B show how policies learned through graph-RL exhibit an interesting degree of portability, substantially outperforming all domain-specific heuristics without having been explicitly trained for transfer performance, resulting in an average improvement of 1.75x.

\smallskip
\noindent \textbf{Service area expansion.} 
To further study how well policies learned through graph-RL can generalize to conditions unseen during training, we now consider the case of a hypothetical service area expansion.
Specifically, we do so by selecting a policy trained within a specific spatial coverage in San Francisco and then deploying it within a larger spatial region (e.g., deploying the policy trained on SF 5 within the SF 20 scenario).
As in the case of inter-city transfer, results in Table \ref{tab:exp1}, Part C highlight how the proposed graph-RL framework exhibits strong intrinsic generalization capabilities and outperforms all domain-specific heuristics, with an improvement ranging between 1.5x and 2.5x of heuristics performance.
Moreover, by comparing results on SF5$\rightarrow$SF20 and NY20$\rightarrow$SF20, our experiments indicate that transfer across cities is more challenging than transfer within the same city: an observation that aligns with intuition, as different cities are typically characterized by more substantial differences (e.g., topology, travel times, etc.).

Together, these experiments highlight the benefits of the inductive biases introduced by graph neural networks and show huge promise in extending these analyses by explicitly considering transfer and generalization in the design of neural architectures and training strategies, e.g., by considering meta-RL\cite{GammelliYangEtAl2022}.

\begin{figure}
  \includegraphics[width=0.95\linewidth]{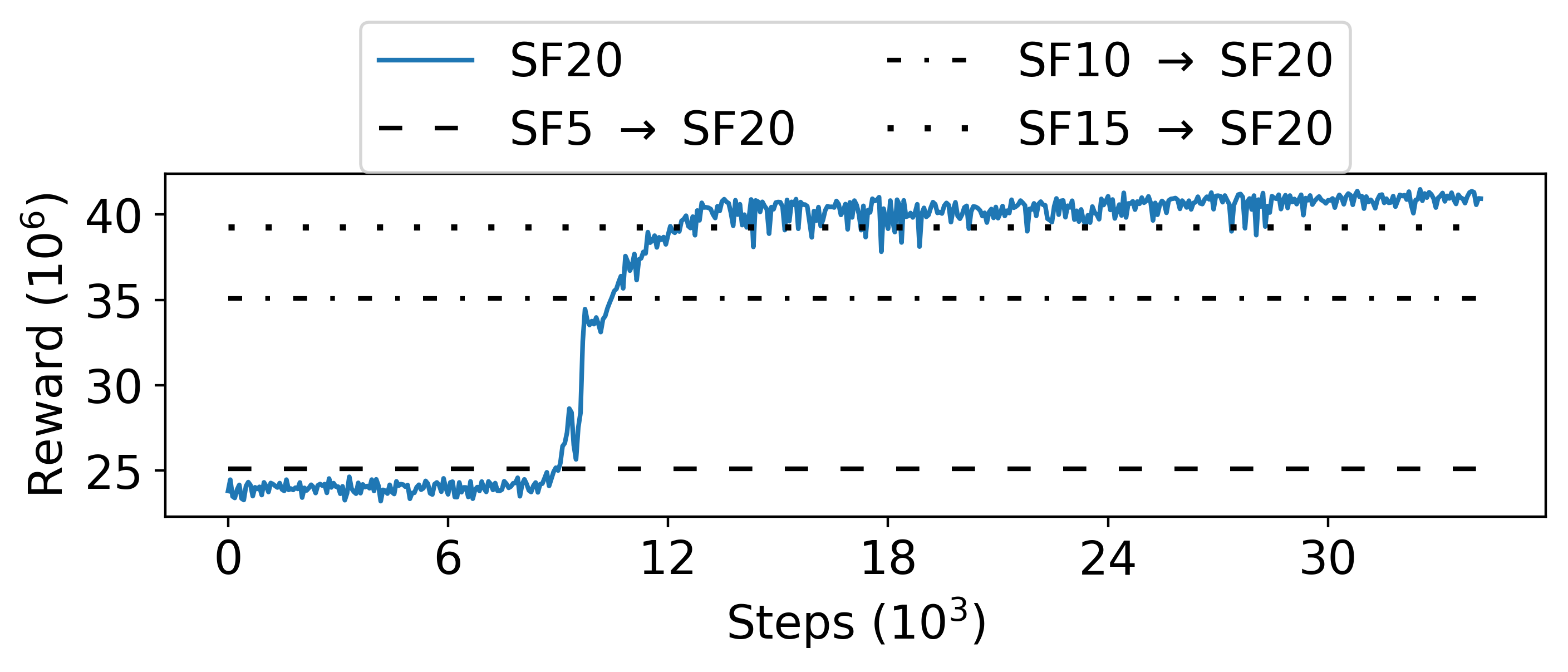}
\caption{Reward curve when training SF20 from scratch compared with zero-shot performance of SF5, SF10, and SF15 agents when deployed on SF20.}
\label{fig:warm-start}
\end{figure}

\smallskip
\noindent \textbf{Transfer to enable learning of large-scale instances.} 
Lastly, we focus on quantifying the potential benefits of the transfer capabilities of graph-RL agents operating within a single city.
Specifically, in Table \ref{tab:exp1}, Part C, and Figure \ref{fig:warm-start}, we measure the zero-shot performance on SF20 of policies trained on smaller scenarios (i.e., SF5, SF10, SF15) as opposed to the performance of training from scratch a new control policy (i.e., SF20).
Results show how the agents trained on SF5, SF10, and SF15 achieve $61.1\%$, $85.4\%$, and $95.5\%$ of the agent fully trained on SF20, respectively. 
Not only does this quantify the benefits of curriculum learning within fleet control problems, whereby more similar environments allow for better transfer, but also opens several promising directions for future work toward the use of agents trained on small, computationally efficient environments as a starting point for successive fine-tuning on larger (and computationally intensive) instances.

\section{CONCLUSIONS}\label{sec:conclusion}

Existing literature on the coordination of E-AMoD systems heavily relies on either optimization-based approaches or domain-specific heuristics.
Among these two classes of techniques, methods belonging to the first have been shown to be extremely performant, although typically not scalable; on the other hand, methods belonging to the second make real-time implementation feasible by sacrificing on performance.
In this paper, we present a graph-RL framework to achieve the best of both worlds: the scalability of heuristics, while maintaining high performance.  
We do so by introducing an RL agent that leverages the benefits of graph neural networks, reinforcement learning, and optimization for the real-time scheduling of E-AMoD fleets.
Our experiments operating an E-AMoD fleet in NYC and SF using realistic data show that 
graph-RL policies can achieve performance comparable to the one of (real-time infeasible) optimization-based approaches while maintaining the scalability of domain-specific heuristics.
Crucially, we show how graph-RL enables reinforcement learning agents to recover highly flexible, generalizable, and scalable behavior policies.
%Our key contribution was to model the E-AMoD setting as a network flow problem using a space-charge-time graph and use a graph RL agent to speed up the computational time required to solve the underlying large-scale optimization. Our experiments on operating an E-AMoD fleet in NYC and SF using realistic data suggest that the RL approach is capable of learning charging and rebalancing policies that beat benchmark heuristics by up to 5\%, and reaches up to 95\% of the cost of an oracle MPC solution with a nearly 200X computational speedup. 
% future directions
There are several avenues for future research. 
First, to further validate the applicability of our method to real-world large-scale E-AMoD systems, a lower-level vehicle dispatch algorithm that is guided by our framework can be integrated with the simulation environment.
Furthermore, the environment can be reconfigured to reward control policies that ensure periodicity in the fleet state for daily repeatability of the operations.
Second, investigating ways to explicitly consider transfer in the design of neural architectures and training strategies, (e.g., meta-learning) is extremely promising.
%While this paper presents a proof of concept that RL can be useful for real-time E-AMoD operations, there is significant scope for evaluating the generalizability of our approach on larger scale problems including those with finer spatial and temporal resolution. Next, we have identified matching as a major factor that determines system performance. Thus there is potential for including an RL agent for the matching problem as well. 
More broadly, the idea that large-scale network control problems can be approximated using a sequence of learning-guided linear approximations that are easier to solve is very promising and merits further exploration.

%\addtolength{\textheight}{-12cm}   % This command serves to balance the column lengths
                                  % on the last page of the document manually. It shortens
                                  % the textheight of the last page by a suitable amount.
                                  % This command does not take effect until the next page
                                  % so it should come on the page before the last. Make
                                  % sure that you do not shorten the textheight too much.

%%%%%%%%%%%%%%%%%%%%%%%%%%%%%%%%%%%%%%%%%%%%%%%%%%%%%%%%%%%%%%%%%%%%%%%%%%%%%%%%

%%%%%%%%%%%%%%%%%%%%%%%%%%%%%%%%%%%%%%%%%%%%%%%%%%%%%%%%%%%%%%%%%%%%%%%%%%%%%%%%

%%%%%%%%%%%%%%%%%%%%%%%%%%%%%%%%%%%%%%%%%%%%%%%%%%%%%%%%%%%%%%%%%%%%%%%%%%%%%%%%

% UNCOMMENT IF NEEDED

% Appendixes should appear before the acknowledgment.
\iftoggle{extended}{
\section*{ACKNOWLEDGMENT}

The authors thank Edward Schmerling, Ruolin Li and James Harrison for their insightful feedback, and Matteo Zallio for help in making the figures.
The authors thank StreetLight Data, Inc. for providing travel demand data and the Stanford Research Computing Center computational resources on the Sherlock cluster.
This research was supported by the National Science Foundation under the CPS program, Stanford Bits \& Watts EV50 Project, the Center for Automotive Research at Stanford, and the NASA University Leadership Initiative (grant \#80NSSC20M0163).
This article solely reflects the opinions and conclusions of its authors and not NSF, NASA, or Stanford University.

}{
% \section*{ACKNOWLEDGMENT}

% The authors thank Edward Schmerling, Ruolin Li and James Harrison for their insightful feedback and suggestions, and Matteo Zallio for help in making the figures. The authors would like to thank StreetLight Data, Inc. for providing travel demand data of San Francisco under the StreetLight Academic Access Agreement and the Stanford Research Computing Center for providing the Sherlock cluster computational resources that contributed to our results. This research was supported by the National Science Foundation under the CPS program, Stanford University Bits \& Watts EV50 Project, the Center for Automotive Research at Stanford, and the NASA University Leadership Initiative (grant \#80NSSC20M0163). This article solely reflects the opinions and conclusions of its authors and not NSF, NASA, or Stanford University.
%
}

%%%%%%%%%%%%%%%%%%%%%%%%%%%%%%%%%%%%%%%%%%%%%%%%%%%%%%%%%%%%%%%%%%%%%%%%%%%%%%%%

\bibliographystyle{IEEEtran}
\bibliography{IEEEexample,ASL_papers}

\end{document}